\documentclass[prb,aps,twocolumn]{revtex4}
\usepackage{epsfig,latexsym,amssymb,amsmath,amsbsy,graphics,graphicx,mathrsfs}
\usepackage[dvips]{feynmf}
\begin{document}
\def \cation{$\kappa$-(ET)$_2$}~
\def \kpx{$\kappa$-(ET)$_2X$}
\def \kbr{$\kappa$-(ET)$_2$Cu[N(CN)$_2$]Br}
\def \deut8br{$\kappa$(d8)-(ET)$_2$Cu[N(CN)$_2$]Br}
\def \h8br{h8-(ET)$_2$Cu[N(CN)$_2$]Br}
\def \kcl{$\kappa$-(ET)$_2$Cu[N(CN)$_2$]Cl}
\def \kncs{$\kappa$-(ET)$_2$Cu(NCS)$_2$}
\def \kcn3{$\kappa$-(ET)$_2$Cu$_2$(CN)$_3$}
\def \>{\textgreater}
\def \<{\textless}
\def \q{\vec{q}}
\def \Q{\vec{Q}}
\def \kpcl{$\kappa$-Cl}
\def \kpbr{$\kappa$-Br}
\def \kpncs{$\kappa$-NCS}
\def \kpcn3{$\kappa$-(CN)$_3$}
\def \d8pbr{$\kappa$(d8)-Br}
\def \m{\mathrm{m}}
\def \max{\mathrm{max}}
\def \cross{\mathrm{cross}}
\def \M{\mathrm{M}}
\def \c{\mathrm{c}}
\def \lw{\mathrm{LW}}
\def \af{\mathrm{AF}}
\def \fm{\mathrm{FM}}
\def \sf{\mathrm{SF}}
\def \res{{\rho \sim T^2}}
\def \us{{\Delta v/v}}
\def \nmr{\mathrm{NMR}}
\def \ks{{K_s}}
\def \etal{{\it et al}. }

\title{Vertex Corrections and the Korringa Ratio in Strongly Correlated Electron Materials}
\author{Eddy Yusuf$^{1,2}$, B. J. Powell$^2$, and Ross H. McKenzie$^2$}
\affiliation{$^1$Department of Physics, University at Buffalo, SUNY,
Buffalo, New York 14260-1500, USA\\$^2$Centre for Organic Photonics
and Electronics, School of Physical Sciences, University of
Queensland, Brisbane, Queensland 4072, Australia}
\date{\today}
\begin{abstract}
We show that the Korringa ratio, associated with nuclear magnetic
resonance in metals, is unity if vertex corrections to the dynamic
spin susceptibility
 are negligible, the   hyperfine coupling
is momentum independent, and there exists an energy scale below which the density of states is constant.
In the absence of vertex corrections we also find a Korringa behaviour for $T_1$, the nuclear spin relaxation rate, i.e., $1/T_1\propto T$, and a temperature independent Knight shift. These results are independent of
the form and magnitude of the self-energy (so far as is consistent with neglecting vertex corrections) and of
the dimensionality of the system.
\end{abstract}

\maketitle

Nuclear
magnetic resonance (NMR) spectroscopy is a powerful
experimental probe of the spin dynamics of strongly correlated
electron materials.
 An important quantity in NMR experiments on metals is the Korringa ratio,\cite{slichter,korringa} ${\cal K}$, which is
proportional to the ratio of the nuclear relaxation rate $1/T_1$ to the
square of the  Knight shift, $K_s$. The
Korringa ratio, and the temperature dependence of $1/T_1$ and $K_s$ can provide important insights into the electronic and magnetic
correlations. 
In this paper we show that, under certain
conditions, the Korringa ratio does not deviate
from its non-interacting value even in strongly correlated electron systems.

In the diagrammatic formalism of quantum many-body theory, the effects of
electronic correlations are described by the self energy and the vertex
corrections. The self energy
describes the effect that interactions with virtual particles have on
the propagation of particles through the material. Vertex corrections describe the
renormalisation of coupling constants due to interactions (the name
arises because coupling constants appear at the vertices in Feynman
diagrams).\cite{mattuck}

An important consideration in the study of vertex corrections is Migdal's theorem, which states
that the vertex corrections due to the electron-phonon interaction are of order $\sqrt{m/M}$, where $m$ is the electron mass and $M$ is the nuclear mass.
The Eliashberg theory of superconductivity, which improves on BCS theory by replacing the BCS effective pairwise interaction with an explicit treatment of the electron-phonon interaction, invokes Migdal's theorem in order to neglect vertex corrections. There is no Migdal's theorem for either the electron-electron interaction or the electron-magnon interaction.\cite{foot-effective}
Therefore, understanding the importance of vertex corrections in strongly correlated superconductors is of great importance and has been widely debated.
 Hertz {\it et al.}\cite{hertz:migdal} have
shown that the first order vertex corrections due to paramagnons is the same order of
magnitude as
the bare vertex. More recently, in the context of the cuprates, arguments have been presented both for\cite{chatterjee,littlewood,varma} and against\cite{millis:eliashberg,schuttler,bang}
 the thesis that vertex corrections are  negligible.  Some theories of strongly correlated superconductors have attempted to deal with the vertex corrections. For example, for the one band Hubbard model, the FLEX approximation\cite{FLEX} consists of a self-consistent summation of bubble and ladder diagrams; the latter are the lowest order self-constient contributions to the vertex function.

The relaxation of the nuclei is governed by their coupling to their
 environment, which, in a metal, is the conduction electrons.\cite{slichter,korringa}
 Hence, many of the properties measured in NMR experiments depend on the transverse dynamic magnetic
 susceptibility of the electron fluid, $\chi_{-+}({\bf q},i\omega_n)$. In Matsubara formalism, this is given by\cite{Doniach&Sondheimer}
\begin{eqnarray}
\chi_{-+}({\bf q},i\omega_n) = \int_0^\beta{d\tau e^{i
\omega_n \tau}\langle T_\tau m_-({\bf q},\tau) m_+(-{\bf
q},0)\rangle}, \label{chi}
\end{eqnarray}
where $\beta=1/k_B T$ is the inverse temperature, $\tau$ is the
imaginary time, $\omega_n$ are the Matsubara
frequencies,
\begin{subequations}\label{ms}
\begin{eqnarray}
m_-({\bf q},\tau) &=& \frac{\hbar\gamma_e}{\sqrt{2}}\sum_{{\bf p}}
c_{{\bf p}+{\bf q},\downarrow}^\dagger(\tau) c_{{\bf
p},\uparrow}(\tau), \label{m} \\
m_+({\bf q},\tau) &=& \frac{\hbar\gamma_e}{\sqrt{2}}\sum_{{\bf p}}
c_{{\bf p}+{\bf q},\uparrow}^\dagger(\tau) c_{{\bf
p},\downarrow}(\tau), \label{mm}
\end{eqnarray}
\end{subequations}
are the $\mp$ components of magnetization,
and $T_\tau$ is the (imaginary) time ordering
operator.

The Korringa ratio\cite{slichter,korringa}  is
the dimensionless quantity
\begin{equation}
\mathcal{K} \equiv \frac{\hbar}{4\pi
k_B}\left(\frac{\gamma_e}{\gamma_N}\right)^2 \frac{1}{T_1T K_s^2},
\label{korringa}
\end{equation}
where
 $\gamma_N$ ($\gamma_e$) is the nuclear (electronic)
gyromagnetic ratio.  The Korringa ratio is unity
 in a non-interacting system
with a contact, i.e. momentum independent, hyperfine coupling. The
hyperfine coupling is momentum independent if there is one atom per
unit cell, and this is an approximation otherwise.\cite{korringa}
Further, Korringa showed that in such a system $1/T_1T$ and $K_s$
are independent of temperature.\cite{korringa} These three
behaviours are often collectively referred to as Korringa behaviour.
However, real materials always exhibit some correlations and the
Korringa ratio may deviate from unity, or $1/T_1T$ or $K_s$ may be
temperature dependent.\cite{moriya:jpsj,narath,shastry:korringa} In
elemental metals the Korringa ratio is typically between 0.6 and 1
(see  Ref. \onlinecite{slichter}, pp. 156-7). Further,  the Korringa
ratio is greater (less) than unity if the system is near an
antiferromagnetic (ferromagnetic) instability.\cite{doniach} There
are large deviations from Korringa behaviour in a wide range of
strongly correlated electron materials including the
cuprates,\cite{mmp} organic charge transfer
salts,\cite{JPCMreview,eddy1} and the heavy fermion
materials.\cite{curro} On the other hand,
 many materials and model Hamiltonians   do show Korringa behaviour;
 including some strongly correlated ones,
such as magnetic impurities
described by the Anderson model.\cite{Hewson}
 Therefore it is
important to determine what conditions may
be sufficient
 for Korringa behaviour in a correlated system.

The relaxation rate and the Knight shift can be written
in terms of the dynamic spin susceptibility,
\begin{subequations}\label{test}
\label{nmr}
\begin{eqnarray}
\frac{1}{T_1T} &=& \lim_{\omega \to 0}\frac{2k_B}{\gamma_e^2\hbar^4}
\sum_{\bf q} |A({\bf q})|^2\frac{\chi_{-+}''({\bf q},\omega)}{\omega},\label{t1t}\\
K_s &=& \frac{|A({\bf 0})| \chi_{-+}'({\bf0},0)}{\gamma_e\gamma_N
\hbar^2}, \label{ks}
\end{eqnarray}
\end{subequations}
where $A({\bf q})$ is the hyperfine coupling between the nuclear and
electron spins, and $\chi'({\bf q},\omega)~[\chi''({\bf
q},\omega)]$ is the real [imaginary] part of the dynamic
susceptibility.

The focus of this paper is on how
electronic correlations affect the Korringa ratio. Specifically, we
investigate how vertex corrections modify the Korringa ratio and what
implications this has for strongly correlated materials. We show
that as zero
temperature
is approached
 the Korringa ratio approaches
unity if three conditions are
satisfied:
(i) vertex corrections are negligible,
 (ii) the hyperfine coupling
 is momentum independent, and (iii) there exits a energy scale below which the density of states is constant.
 This result holds independent of the strength
of the electron-electron interactions that enter the
 self energy and
of the dimensionality of
the system.
 We also find that, under the same conditions, $1/T_1T$ and $K_s$ are independent of temperature.
Hence,  non-Korringa behaviour must result from vertex corrections, a momentum dependent hyperfine coupling and/or the absence of an energy scale below which the density of states is constant.

Upon substituting (\ref{ms}) into (\ref{chi}),
performing the appropriate Wick contractions on the operators, and Fourier transforming  into frequency
space one finds that
\begin{eqnarray}
\nonumber\\
\chi_{-+}({\bf q},i\omega_n) &=&
\begin{fmffile}{chi}
        \begin{fmfgraph*}(20,1)
            \fmfleft{i}
            \fmfright{o}
            \fmf{phantom}{i,G3}
            \fmfpoly{shaded,tension=0.01,label=$\bf\Gamma$}{G3,G2,G1}
            \fmf{fermion,left=0.3,tension=0.01,label=${\bf p} \uparrow$}{G1,o}
            \fmf{fermion,left=0.3,tension=0.01,label=${\bf p}+{\bf q} \downarrow$}{o,G2}
        \end{fmfgraph*}
    \end{fmffile}
    \\ \nonumber
    \\ \nonumber
    \\&=&
\frac{\hbar^2\gamma_e^2}{2\beta}\sum_{{\bf p},ip_m}\Gamma({\bf p}+{\bf
q},ip_m;{\bf p},ip_m+i\omega_n)
\nonumber\\&&\times
G({\bf p}+{\bf
q},ip_m)G({\bf p},ip_m+i\omega_n)\label{chi_zz_2},
\end{eqnarray}
where $\Gamma({\bf q},i\omega_n;{\bf p},i\omega_n')$ [shaded area] is the three point vertex function and $G({\bf
p},ip_n)$ [solid lines] is the full interacting Green's function given, in the spectral representation, by
\begin{equation}
G({\bf p},ip_n) =
\int_{-\infty}^{\infty}{\frac{dE_1}{2\pi}\frac{A_s({\bf
p},E_1)}{ip_n-E_1}}, \label{lehmann}
\end{equation}
where $A_s({\bf p},E_1)$ is the spectral function, given by
\begin{eqnarray}
A_s({\bf p},E) = \frac{-2\mathrm{Im}\Sigma({\bf p},E)}
{(E-\varepsilon_{\bf p}-\mathrm{Re}\Sigma({\bf
p},E))^2+(\mathrm{Im}\Sigma({\bf p},E))^2}, \label{spectral}
\end{eqnarray}
 $\varepsilon_{\bf p}$ is the dispersion of the non-interacting
system, and
$\Sigma({\bf p},ip_n)$ is the self energy.
Therefore
\begin{eqnarray}
\chi_{-+}({\bf q},i\omega_n) &=&
\frac{\hbar^2\gamma_e^2}{2\beta}\sum_{{\bf p},m}\int_{-\infty}^{\infty}
\frac{dE_1}{2\pi}\frac{dE_2}{2\pi}\nonumber\\
&&\times\nonumber\Gamma({\bf p}+{\bf q},ip_m;{\bf p},ip_m+i\omega_n)
\\&&\times\frac{A_s({\bf p}+{\bf q},E_1)A_s({\bf
p},E_2)}{(ip_m-E_1)(ip_m+i\omega_n-E_2)}.
\end{eqnarray}

At this stage we neglect vertex corrections, that is we set
$\Gamma({\bf q},i\omega_n;{\bf p},ip_n)=1$ for all ${\bf
p}$, ${\bf q}$, $p_n$, and $\omega_n$. After performing the Matsubara
summation
and the analytical continuation $i\omega_n \to \omega + i\eta$, one finds that
\begin{eqnarray}
\chi_{-+}({\bf q},\omega) &=& \frac{\hbar^2\gamma_e^2}{2} \sum_{{\bf
p}}\int_{-\infty}^{\infty}\frac{dE_1}{2\pi}\frac{dE_2}{2\pi}A_s({\bf
p}+{\bf q},E_1)
\nonumber\\
&& \times A_s({\bf p},E_2)
\frac{n_F(E_1)-n_F(E_2)}{\hbar\omega+E_1-E_2+i\eta},\label{chi_final}
\end{eqnarray}
where $n_F(E)$ is the Fermi function.

Using the well known equality $1/(x+i\eta) = {\cal P}(1/x) - i \pi
\delta(x)$, where ${\cal P}(y)$ denotes the principal value, one finds
\begin{eqnarray}
\lim_{\omega \to 0} \frac{\chi''_{-+}({\bf q},\omega)}{\hbar\omega}
&=& \frac{\hbar^2\gamma_e^2}{2}\sum_{{\bf p}}
\int_{-\infty}^{\infty}\frac{dE}{4\pi}A_s({\bf p}+{\bf q},E)
\nonumber\\ && \hspace{1.1cm}\times
A_s({\bf
p},E) \left(-\frac{\partial n_F}{\partial
E}\right). \label{im_chi_final}
\end{eqnarray}
It then follows from Eq. (\ref{t1t}) that
 \begin{eqnarray}
\frac1{T_1T} = \frac{k_B |A|^2}{\hbar}
\int_{-\infty}^{\infty}\frac{dE}{4\pi}\tilde{\rho}^2(E)
\left(-\frac{\partial n_F}{\partial E}\right),\label{t1t_local1}
\label{rate}
\end{eqnarray}
where
$\tilde{\rho}(E)=\sum_{\bf p} A_s({\bf p},E)$ is the full interacting density of states per spin species and we have assumed a contact hyperfine coupling, $A({\bf q})=A$ for all ${\bf q}$. We now specialise to the case where their exists an energy scale, $k_B T_0$ below which $\tilde{\rho}(E)$ is independent of energy. There several situations in which $T_0$ may not exist or may be too small to be of practical interest, for example, if the Fermi energy is at, or very close to, a van Hove singularity or if there is a (pseudo)gap at the Fermi energy.
For $T\ll  T_0, T_F$, where  $T_F$ is the Fermi temperature,
Eq. (\ref{rate}) simplifies to
\begin{equation}
\frac1{T_1T} \simeq \frac{k_B |A|^2
\tilde{\rho}^2(E_F)}{4\pi\hbar}.\label{t1t_local2}
\end{equation}
Note that the right hand side, and therefore $1/T_1T$, is independent of temperature.


For $H\rightarrow0$, where $H$ is a static magnetic field,
$\chi_{-+}^\prime({\bf0},0)=\chi_{zz}^\prime({\bf0},0)=\partial M/\partial
H|_{H=0}$, where $\chi_{zz}^\prime({\bf q},\omega)$ is the real part of the longitudinal
component of the dynamic spin susceptibility and the magnetisation,
$M$ is given by\cite{Luttinger}
\begin{eqnarray}
M&=&\frac{\hbar\gamma_e}{2}\sum_{\bf k}\big(\langle \hat n_{{\bf
k}\uparrow}\rangle-\langle \hat n_{{\bf k}\downarrow}\rangle\big),
\end{eqnarray}
where $\hat n_{{\bf k}\sigma}$ is the usual number operator. Writing
$\hat n_{{\bf k}\sigma}$ in the spectral representation and
performing the sum over ${\bf k}$ one finds that
\begin{eqnarray}
M = \frac{\hbar\gamma_e}{2} \sum_{\sigma} \sigma
\int_{-\infty}^\infty \frac{dE}{2\pi} \tilde\rho_{\sigma}({\bf k},E)
n_{F\sigma}(E),
\end{eqnarray}
where we take $\sigma=\pm1$, $\rho_\sigma(E)=\sum_{\bf
k}A_{s\sigma}({\bf k},E)$ is the density of states of spin $\sigma$
electrons and $A_{s\sigma}({\bf k},E)$ is the spectral function for
spin $\sigma$ electrons. Even for small magnetic fields the Fermi
surface may, in general, be distorted\cite{Luttinger} as $E_{{\bf
k}\sigma}=E_{{\bf k}}-\sigma H\Gamma({\bf k},E_{{\bf k}\sigma};{\bf
k},E_{{\bf k}\sigma})\hbar\gamma_e/2+{\cal O}(H^2)$. However, in the
absence of vertex corrections such complications cannot arise as the
$\bf k$-dependence drops out of the above equation. Therefore we
deal with the magnetic field by introducing a spin dependent
chemical potential: $\mu_{\sigma}=\mu-\sigma H\hbar\gamma_e/2+{\cal
O}(H^2)$ and noting that $n_{F\sigma}=\{1+\exp\beta[(E-\mu_\sigma)]\}^{-1}$. Thus one finds that,
\begin{widetext}
\begin{eqnarray}
\lim_{T\rightarrow0} M &=& \frac{\hbar\gamma_e}{2} \sum_{{\bf
k}\sigma} \sigma \int_{-\infty}^\infty \frac{dE}{2\pi} \left[
\tilde\rho_{\sigma}(E) \Theta(E-\mu_\sigma) + \sigma \frac{\hbar
\gamma_e}{2} \tilde\rho_{\sigma}(E) \delta(E-\mu_\sigma) H +
\frac{\partial \tilde\rho_{\sigma}(E)}{\partial H}
\Theta(E-\mu_\sigma) +{\cal O}(H^2) \right].
\end{eqnarray}
In the limit $H\rightarrow0$ the first term vanishes due to spin symmetry and the third term vanishes from the assumption that we are at an energy scale on which the density of states is constant. Thus, the Knight shift  is 
\begin{equation}
K_s \simeq
\frac{|A|\gamma_e\tilde{\rho}(E_F)}{4\pi\gamma_N},\label{ks_local2}
\end{equation}
which is independent of temperature.

It follows immediately from Eqs. (\ref{t1t_local2}) and (\ref{ks_local2}), that $\mathcal{K}=1$ for interacting electrons with a contact hyperfine coupling when $T\ll T_0,T_F$ and
neglecting vertex corrections.
We stress two points about this result:
firstly, our result includes the
special case of Korringa's result for
the free electron gas,
${\cal K}_\mathrm{free}=1$;
secondly, and more importantly, the Korringa
ratio is unity for a broad class
of systems and not just for the free electron gas.

Thus we have shown that any deviation of the
Korringa ratio from unity, or temperature dependence of $1/T_1T$ or $K_s$ must be caused by either vertex corrections,
 the  wavevector
dependence of the hyperfine coupling, or the fact the there is no energy, $k_BT_0$, on which we may treat the density of states as constant.
 Note that this result is independent of the dimensionality of the system.
Further, this is true for any form of the self energy.
However, it should be
noted that the self-energy and the vertex function are not really independent: they both arise from the same underlying
 interactions and can
often be related by Ward identities.
 So this last statement should be taken with appropriate
caution.

Let us now briefly discuss the role of vertex corrections in  strongly correlated systems. A simple and
a widely studied approach to nearly (anti)ferromagetic metals is the random phase approximation (RPA).
 For the Hubbard model the RPA
 gives,\cite{Doniach&Sondheimer}
\begin{eqnarray}
\chi_\mathrm{RPA}({\bf q},\omega)&=&
\begin{fmffile}{rpat}
        \begin{fmfchar*}(20,9)
            \fmfleft{i}
            \fmfright{o}
                \fmftop{v1}
            \fmfbottom{v2}
            \fmf{phantom,left=0.4,label=$\uparrow$}{i,o}
            \fmf{phantom,left=0.4,label=$\downarrow$}{o,i}
               \fmffreeze
            \fmf{phantom}{i,G3}
            \fmfpoly{shaded,tension=0.01,label=$\hspace{0.5cm}\bf\Gamma\hspace{-0.08cm}_\textrm{RPA}$}{G3,G2,G1}
            \fmf{fermion,left=0.3,tension=0.01}{G1,o}
            \fmf{fermion,left=0.3,tension=0.01}{o,G2}
        \end{fmfchar*}
    \end{fmffile}
=
        \begin{fmffile}{rpa0}
        \begin{fmfchar*}(20,9)
            \fmfleft{i}
            \fmfright{o}
                \fmftop{v1}
            \fmfbottom{v2}
            \fmf{fermion,left=0.4,label=$\uparrow$}{i,o}
            \fmf{fermion,left=0.4,label=$\downarrow$}{o,i}
        \end{fmfchar*}
    \end{fmffile}+
    \begin{fmffile}{rpa1}
        \begin{fmfchar*}(20,9)
            \fmfleft{i}
            \fmfright{o}
                \fmftop{v1}
            \fmfbottom{v2}
                \fmf{fermion,left=0.2}{i,v1,o}
            \fmf{fermion,left=0.2}{o,v2,i}
            \fmf{dashes,label=$U$,l.d=.03w}{v1,v2}
               \fmflabel{$\uparrow$}{v1}
               \fmflabel{$\downarrow$}{v2}
        \end{fmfchar*}
    \end{fmffile}
    +
    \begin{fmffile}{rpa2}
        \begin{fmfchar*}(20,9)
            \fmfleft{i}
            \fmfright{o}
                \fmftop{x1,w1,v1,u1}
            \fmfbottom{x2,w2,v2,u2}
                \fmf{fermion,left=0.1,label=$\uparrow$}{w1,v1}
            \fmf{fermion,left=0.1,label=$\downarrow$}{v2,w2}
                \fmf{fermion,left=0.2}{i,w1}
            \fmf{fermion,left=0.2}{o,v2}
                \fmf{fermion,left=0.2}{v1,o}
            \fmf{fermion,left=0.2}{w2,i}
            \fmf{dashes,label=$U$,l.d=.03w}{v1,v2}
            \fmf{dashes,label=$U$,l.d=.03w}{w1,w2}
        \end{fmfchar*}
    \end{fmffile}+\dots
=
    \frac{\chi_0({\bf q},\omega)}{1-U\chi_0({\bf q},\omega)},
\label{eqn:chi-rpa}
\end{eqnarray}
\end{widetext}
where $U$ (dashed lines) is the effective Coulomb interaction between electrons on the same lattice site, $\chi_0$ is the dynamic susceptibility in the {\it absence} of vertex corrections, and we have suppressed momentum labels in the diagrams for clarity. The distinctive `Stoner-like' form of Eq.~(\ref{eqn:chi-rpa})
 arises from the sum of ladder diagrams, i.e., vertex corrections.
For this form of $\chi({\bf q},\omega)$
and a three-dimensional parabolic band the
Korringa ratio is always less than unity and decreases monotonically
towards zero as the Stoner instability is approached.\cite{narath}
However, for
a two-dimensional parabolic band the free electron spin susceptibility is momentum
independent. Consequently, c.f. Eqs. (\ref{test}), the Korringa ratio remains
unity for all $U$ even though $1/T_1$ and the Knight shift
both diverge as the Stoner instability is approached.\cite{nishihara:korringa}

Korringa type relations have been derived
for
the impurity spin susceptibility,
$\chi_{imp}(\omega)$,
of the
Anderson single impurity model.
It is found that\cite{zawadowski,Hewson}
\begin{equation}
\lim_{\omega \to 0}
\frac{\chi_{imp}''(\omega)}{\omega} =
 \frac{2 \pi \chi_{imp}'(0)^2}{\gamma_e^2},
\label{shiba}
\end{equation}
which is sometimes referred to as the Shiba relation.
This relation holds even though there can be
significant vertex corrections for the spin susceptibility.
This and the case of the two-dimensional
RPA with a parabolic band illustrate that
{\it the absence of vertex corrections
is sufficient but not necessary for the Korringa ratio to be unity.}

 A similar form of $\chi({\bf q},\omega)$
to that given in (\ref{eqn:chi-rpa})
 is found in dynamical mean field theory (DMFT),
 but with $U$ replaced by a self-consistently determined four-point vertex function.\cite{Georges}
 In DMFT, as in the RPA, this functional form results from a sum over ladder diagrams, with the four-point vertex function now forming the legs of the ladders.
DMFT for the Hubbard model
is equivalent to an Anderson single impurity model
which is solved self-consistently.\cite{Georges}
It follows that the local spin susceptibility
$\chi_{loc}(\omega) = \sum_{\bf q} \chi({\bf q},\omega)$ must satisfy the Shiba relation (\ref{shiba}) with
$\chi_{imp}(\omega) = \chi_{loc}(\omega).$
Since, in general,
$\chi_{loc}(\omega) \neq  \chi({\bf q}=0,\omega)$
the Korringa relation does not necessarily hold
in DMFT [see Eq. (73) in Ref. \onlinecite{Georges}].

A slave-boson large $N$ treatment of the periodic Anderson
model shows that in the low-temperature heavy fermion
phase the Korringa ratio is close to unity,
even though $1/T_1T$ and $K_s$ can be
several orders of magnitude larger than the
value predicted in the
 absence of strong electronic correlations.\cite{evans}

In the singlet superconducting states the opening of the gap causes a suppression of both $1/T_1T$ and $K_s$. For a fully gapped superconductor both $1/T_1T$ and $K_s$ show activated behaviours, while if the order parameter has nodes power laws are seen at low $T$.\cite{M&S} The behaviour, particularly of the Knight shift, is more complicated in a triplet superconductor.\cite{M&S} In the same way, a (pseudo)gap destroys the energy scale $T_0$ and hence causes non-Korringa behaviour by \emph{suppressing} $1/T_1T$ and $K_s$.

Given the above discussion it is interesting to consider the normal states of a few superconductors which display non-Korringa behaviour.
 A strong temperature dependence and a  Korringa
ratio significantly larger than unity
is observed in the organic charge transfer salts,\cite{JPCMreview,eddy1} e.g., in
\kbr~(Refs. \onlinecite{desoto,miyagawa}). Heavy fermion
compounds\cite{buttgen} and cobaltates\cite{vaulx}
also show enhanced Korringa ratios above their
Kondo temperatures. Most recently non-Korringa behaviour has been observed in the iron pnictides.\cite{Ishida}
 In the cuprates, the Korringa ratio of
YBa$_2$Cu$_3$O$_{6.64}$ (Refs. \onlinecite{takigawa,walstedt}) has a
strong temperature dependences and is larger than unity. However, in
the cuprates the hyperfine coupling has a significant wave vector
dependence.\cite{mila2,shastry} This complicates deducing the
relative importance of vertex corrections. We also note that vertex
corrections are required in the superconducting state in order to
preserve gauge invariance.\cite{Schrieffer}

  Moriya's self-consistent renormalised theory\cite{moriya:jpsj} and the phenomenological MMP theory\cite{mmp}
 give a good description of many features
of the cuprates,\cite{mmp,moriya,curro} organics,\cite{eddy1} and heavy fermion materials,\cite{curro}. These theories
 posit a form of $\chi({\bf q},\omega)$ which
follows from the form of Eq.~(\ref{eqn:chi-rpa}).
Hence, these theories  implicitly include vertex corrections.

In summary, we have shown that  for a system with a contact hyperfine coupling the Korringa ratio is unity in the absence of vertex
corrections provided there is an energy scale on which the density of states may be treated as constant. At
sufficiently low temperatures
$1/T_1T$ and the Knight shift are both independent of temperature under the same assumptions. 

This work was supported by the Australian Research Council (ARC)
under the Discovery scheme (project DP0557532).
BJP  was the recipient of  an ARC Queen
Elizabeth II Fellowship (DP0878523). RHM  was the recipient of  an ARC Professorial Fellowship (DP0877875).


\end{document}